\documentstyle[aps,12pt]{revtex}
\begin{document}
\draft
\author{O. B. Zaslavskii}
\address{Department of Physics, Kharkov Karazin's National University, Svobody Sq.4,
Kharkov\\
61077, Ukraine\\
E-mail: aptm@kharkov.ua}
\title{Quantum extreme black holes at finite temperature and exactly solvable
models of 2d dilaton gravity}
\maketitle

\begin{abstract}
It is argued that in certain 2d dilaton gravity theories there exist
self-consistent solutions of field equations with quantum terms which
describe extreme black holes at nonzero temperature. The curvature remains
finite on the horizon due to cancelation of thermal divergencies in the
stress-energy tensor against divergencies in the classical part of field
equations. The extreme black hole solutions under discussion are due to
quantum effects only and do not have classical counterparts.
\end{abstract}

\pacs{PACS numbers: 04.60Kz, 04.70.Dy}

\section{Introduction}

In \cite{hawking} it was argued that thermodynamic reasonings allow one to
ascribe nonzero temperature $T$ to extreme black holes since the Euclidean
geometry remains regular irrespectively of the value of $T$. As a result, it
was conjectured that an extreme black hole may be in thermal equilibrium
with ambient radiation at any temperature. This conclusion was criticized in 
\cite{anders} where it was pointed out that the prescription of \cite
{hawking} is suitable only in the zero-loop approximation and does not
survive at quantum level since the allowance for quantum backreaction leads
to thermal divergencies in the stress-energy tensor that seems to destroy a
regular horizon completely, so the demand of regularity of this tensor on
the horizon seems to enforce the choice $T=0$ for extreme black holes
unambiguously.

The aim of the present paper is to show that in dilaton gravity there exists
possibility to combine a finite curvature at the horizon with divergencies
of the stress-energy tensor $T_{\mu }^{\nu }$. As a result, extreme black
holes with $T\neq 0$ and finite curvature on the horizon may exist. Namely,
the above mentioned divergencies may under certain conditions be compensated
by the corresponding divergencies in the classical part of field equations
due to derivatives of the dilaton field. We stress that the geometries
discussed below are self-consistent solutions of field equations with
backreaction of quantum fields taken into account. In spite of the geometry
itself turns out to be regular in the sense that the curvature measured from
outside is finite at the horizon, the solution as the whole which includes,
apart from the metric, the dilaton field as well, is singular. (The similar
result was obtained quite recently for nonextreme black holes \cite{non}).

It was observed earlier that ''standard'' extreme black holes with $T=0$
exhibit some weak divergencies at the horizon \cite{triv}. The phenomenon we
are discussing is qualitatively different. First, these divergencies are
inherent in our case to the components $T_{\mu }^{\nu }$ themselves, whereas
in \cite{triv} $T_{\mu }^{\nu }$ were assumed to be finite and divergencies
revealed themselves in the energy measured by a free falling observer which
is proportional to the ratio $(T_{1}^{1}-T_{0}^{0})/g_{00\text{ }}$($x^{0}$
and $x^{1}$are a temporal and spatial coordinates in the Schwarzschild
gauge). Second, divergencies considered in \cite{triv} arise for a generic
extreme black hole with $T=0$, whereas in our case they are due to $T\neq 0$
entirely. Third, backreaction of quantum field produces a singularity at the
horizon in \cite{triv}, whereas in our case the curvature remains finite
there.

It is worth noting that in quantum domain the issue of existence of extreme
black holes is non-trivial by itself. In Ref. \cite{extnon} it was argued
that the existence of quantum extreme black holes is inconsistent with field
equations at all. However, this conclusion was derived for the particular
class of CGHS-like models \cite{callan}, so its region of validity is very
limited, as, in fact, the authors themselves note at the end of their paper.
In the present paper we show that for a certain class of dilaton gravity
theories (i) extreme black holes with quantum corrections taken into account
do exist; (ii) the temperature of quantum fields in their background may be
nonzero. We exploit the approach which was elaborated earlier in Refs. \cite
{zasl99}, \cite{throat}, where it was applied to nonextreme black holes and
semi-infinite throats.

\section{General structure of solutions}

Consider the action of the dilaton gravity 
\begin{equation}
I=I_{0}+I_{PL}\text{,}  \label{action}
\end{equation}
where 
\begin{equation}
I_{0}=\frac{1}{2\pi }\int_{M}d^{2}x\sqrt{-g}[F(\phi )R+V(\phi )(\nabla \phi
)^{2}+U(\phi )]  \label{clac}
\end{equation}
and the Polyakov-Liouville action \cite{pl} incorporating effects of Hawking
radiation and its backreaction on the black hole metric can be written as 
\begin{equation}
I_{PL}=-\frac{\kappa }{2\pi }\int_{M}d^{2}x\sqrt{-g}[\frac{(\nabla \psi )^{2}%
}{2}+\psi R]\text{.}  \label{PL}
\end{equation}
The function $\psi $ obeys the equation 
\begin{equation}
\Box \psi =R\text{,}  \label{psai}
\end{equation}
where $\Box =\nabla _{\mu }\nabla ^{\mu }$, $\kappa =N/24$ is the quantum
coupling parameter, $N$ is number of scalar massless fields, $R$ is a
Riemann curvature. We omit the boundary terms in the action as we are
interested only in field equations and their solutions.

A generic quantum dilaton-gravity system is not integrable. However, if the
action coefficients obey the relationship 
\begin{equation}
V=\omega (u-\frac{\kappa \omega }{2})\text{,}  \label{exact}
\end{equation}
where $\omega =U^{\prime }/U$, the system becomes exactly solvable \cite{kaz}%
, \cite{zasl99}. Then static solution are found explicitly and even for a
finite arbitrary temperature $T$ of quantum fields measured at infinity \cite
{throat}: 
\begin{eqnarray}
ds^{2} &=&g(-dt^{2}+d\sigma ^{2})\text{, }g=\exp (-\psi _{0}+2y)\text{, }%
y=\lambda \sigma \text{, }\psi _{0}=\int \omega d\phi =\ln U(\phi )+const%
\text{.}  \label{bas} \\
F^{(0)}(\phi ) &=&f(y)\equiv e^{2y}-By+C\text{, }B=\kappa (1-T^{2}/T_{0}^{2})%
\text{, }T_{0}=\lambda /2\pi \text{,}  \nonumber
\end{eqnarray}
and the auxiliary function $\psi =\psi _{0}+\gamma \sigma $, $\gamma
=2\lambda (T/T_{0}-1)$. Here $F^{(0)}=F-\kappa \psi _{0}$, $y=\lambda \sigma 
$, $\lambda =\sqrt{U}/2$. At right infinity, where spacetime is supposed to
be flat, the stress-energy tensor has the form 
\begin{mathletters}
\begin{equation}
T_{\mu }^{\nu (PL)}=\frac{\pi N}{6}T^{2}(1,-1)\text{.}  \label{T}
\end{equation}
The coordinate $\sigma $ is related to the Schwarzschild one $x$, where 
\end{mathletters}
\begin{equation}
ds^{2}=-dt^{2}g+g^{-1}dx^{2}\text{,}  \label{sch}
\end{equation}
by the formula $x=\int gd\sigma $. We assume that the function $\omega (\phi
)$ changes its sign nowhere, so in what follows we may safely use the
quantity $\psi _{0}$ instead of $\phi $ (in fact, this can be considered as
reparametrization of the dilaton).

The solutions (\ref{bas}) include different types of object - nonextreme
black holes, ''semi-infinite throats'' and soliton-like solutions, depending
on the boundary conditions imposed on the function $\psi _{0}$ on the
horizon \cite{throat}. Now we want to elucidate whether the models (\ref
{exact}) include also extreme black holes with a finite curvature at the
horizon and under what conditions. Near a horizon the metric of a typical
extreme black hole must have the standard form (subscript ''h'' indicates
that a quantity is calculated at the horizon) 
\begin{equation}
g\approx g_{0}(x-x_{h})^{2}\approx \frac{g_{0}}{y^{2}}  \label{g}
\end{equation}
or, equivalently, 
\begin{equation}
\psi _{0}\approx 2y+\ln y^{2}  \label{psi}
\end{equation}
At the right infinity the function $\psi _{0}$ must obey the relation 
\begin{equation}
\psi _{0}\approx 2y  \label{inf}
\end{equation}
that a spacetime have the Minkowski form and, thus, $\psi _{0}\rightarrow
+\infty $ in accordance with general properties indicated in \cite{sol95}.

Provided eqs. (\ref{g}) and (\ref{psi}) are satisfied, the Hawking
temperature 
\begin{equation}
T_{H}=\frac{1}{4\pi }(\frac{dg}{dx})_{h}=\lim_{y\rightarrow -\infty }\frac{%
\lambda dg}{4\pi gdy}=\frac{\lambda }{2\pi }\lim_{y\rightarrow -\infty }(1-%
\frac{1}{2}\frac{d\psi _{0}}{dy})=0  \label{htemp}
\end{equation}
as it should be for an extreme black hole, the Riemann curvature $R=-\lambda
^{2}g^{-1}\frac{d^{2}}{dy^{2}}\ln g\approx -2g_{0}^{-1}\lambda ^{2}$is
finite. As near the horizon $y\rightarrow -\infty $, $f(y)\approx -By$, we
obtain that the function 
\begin{equation}
F^{(0)}\approx -\frac{B}{2}(\psi _{0}-\ln \frac{\psi _{0}^{2}}{4})  \label{F}
\end{equation}
at $\psi \rightarrow -\infty $.

If eq. (\ref{F}) is satisfied, the function $\psi _{0}(\phi )$ which is the
solution of (\ref{bas}) has the desirable asymptotic (\ref{psi}) and,
therefore, the metric function behaves according to (\ref{g}). To achieve
the behavior (\ref{inf}) at infinity, it is sufficient to enforce the
condition $F^{(0)}\approx e^{\psi _{0}}$ at $\psi \rightarrow +\infty $.

As we are looking for solutions of field equations which are regular in the
region between the horizon and infinity, we must exclude a possible
singularity of curvature. Since $R=-\lambda ^{2}g^{-1}\frac{d^{2}\psi _{0}}{%
dy^{2}}$ and $\frac{d\psi _{0}}{dy}=\frac{f^{\prime }(y)}{F^{\prime }(\psi
_{0})}$ (prime denotes derivative with respect to a corresponding argument),
the dangerous point is $\psi _{0}^{*}$ where $F^{(0)\prime }(\psi
_{0}^{*})=0 $ and $\frac{d\psi _{0}}{dy}$ may diverges. Let $B>0$. Then $%
\frac{d\psi _{0}}{dy}$ is finite in the corresponding point $y^{*}$,
provided $f^{\prime }(y^{*})=0$. It is achieved by the choice $%
C=F^{(0)}(\psi _{0}^{*})+\frac{B}{2}(\ln \frac{B}{2}-1)\equiv C^{*}$. If $%
B<0 $, the function $F^{(0)}(\psi _{0})$ must be monotonic to ensure the
absence of singularities. For $B=0$, as we will see below, extreme black
holes do not exist.

Let us consider an example for which all above-mentioned conditions are
satisfied: 
\begin{equation}
F^{(0)}=e^{2\gamma }-B\gamma +C^{*}\text{,}  \label{ex}
\end{equation}
where $\gamma =\gamma (\psi _{0})$. Then eq. (\ref{bas}) has the obvious
solution: $y=\gamma (\psi _{0})$. If we choose the function $\gamma $ in
such a way that the solution of this equation $\psi _{0}(y)$ has the needed
asymptotics (\ref{psi}), we obtain an extreme black hole in accordance with (%
\ref{g}). It is also clear that $F^{(0)\prime }=0$ in the same point where $%
f^{\prime }(y)=0$, so the condition of absence of singularities is satisfied.

Let us look now at the behavior of the components of the stress-energy
tensor of quantum field near the horizon. For definiteness, let us choose
the $T_{1}^{1}$ component. It can be written in the Schwarzschild gauge as
(see, for example, \cite{anders}; it is assumed that the field is
conformally invariant, our definition of $T_{\mu }^{\nu (PL)}$ differs by
sign from the quantum stresses discussed in \cite{anders}) 
\begin{mathletters}
\begin{equation}
T_{1}^{1(PL)}=\frac{\pi }{6g}[T^{2}-(\frac{g^{\prime }}{4\pi })^{2}]\text{.}
\label{stress}
\end{equation}
Since in our case $T\neq T_{H}$, the expression in square brackets remains
nonzero at the horizon, $T_{1}^{1}\sim g^{-1}\sim (x-x_{h})^{-2}\sim y^{2}$,
so stresses diverge strongly.

\section{properties of solutions}

It was observed in \cite{zasl99} that for black holes described exact
solutions (\ref{bas}) the Hawking temperature $T_{H}=\lambda /2\pi $ is
nonzero constant irrespectively of the particular kind of the model that
generalizes the earlier observation \cite{solod} for the RST\ model \cite
{rst}. Therefore, it would seem that black holes entering the class of
solutions under discussion may be nonextreme only. The reason why, along
with nonextreme holes, the class of solutions contains, nevertheless,
extreme ones too, can be explained in terms of the function $\psi $. It was
assumed in \cite{solod} and \cite{zasl99} that this quantity is finite on
the horizon to ensure the finiteness of the Riemann curvature. Then the
derivative $\frac{d\psi }{dy}\rightarrow 0$ at the horizon, where $%
y\rightarrow -\infty $, and $T_{H}=\lambda /2\pi $ in accordance with (\ref
{htemp}). However, we saw that even notwithstanding $\psi $ diverges on the
horizon, we can find the solutions with a finite $R$ on the horizon,
provided the function in question has the asymptotic (\ref{psi}). Thus, we
gain qualitatively new types of solutions due to relaxing the condition of
the regularity of $\psi $ on the horizon.

It is worth paying attention to two nontrivial features of solutions under
discussion:\ (i) the character of solutions in the classical ($\kappa =0$)
and quantum ($\kappa \neq 0$) cases is qualitatively different, (ii) extreme
black holes with a finite curvature at the horizon are possible even if $%
T\neq T_{H}=0$. First, let us consider the point (i). Even if $T=0$, in the
quantum case the coefficient $B\neq 0$ and $f\rightarrow \infty $ at the
horizon, whereas in the classical case ($B=0$) $f\rightarrow C=const$.
Therefore, in the classical case the value of the coupling coefficient $F$
is finite on the horizon: according to (\ref{bas}), $F_{h}=F_{h}^{(0)}=C$.
As a result, in the vicinity of the horizon $\psi _{0}=\psi _{0h}$ plus
exponentially small corrections, so $\psi _{0}(y)$ is regular and a
nonextreme black hole becomes nonextreme, as explained in the precedent
paragraph. On the other hand, $F^{(0)}$ must diverge at the horizon in the
quantum case. Thus, we arrive at a rather unexpected conclusion: for exactly
solvable models of dilaton gravity (\ref{exact}) extreme black holes are due
to quantum effects only and disappear in the classical case $\kappa =0$.
This fact is insensitive to the value of temperature, so classical extreme
black holes of the given type are absent even if $T=0$. For the same reasons
quantum extreme black hole are absent in the exceptional case $T=T_{0}$,
when $B=0$.

It was observed in \cite{extnon} that for a rather wide class of Lagrangians
the typical situation is such that classical extreme black holes may exist,
whereas quantum correction to the action destroy the character of the
solution completely and do not allow the existence of extreme black holes.
In our case, however, the situation is completely opposite: extreme black
holes are possible in the quantum (but not classical) case. It goes without
saying that our result does not contradict the existence of extreme black
holes in the classical dilaton gravity theories in general but concerns only
the special class of Lagrangians (\ref{exact}) and their classical limit.
(It is worth recalling that the condition (\ref{exact}) was imposed to
ensure the solvability of the quantum theory, whereas in the classical case
every dilaton-gravity model is integrable, so there is no need in
supplementary conditions like (\ref{exact})). The results of either \cite
{extnon} or the present paper show clearly that quantum effects not only may
lead to quantum corrections of the classical metric but radically change the
character of the geometry and topology.

The most interesting feature of solutions obtained in the present paper is
the possibility to have black holes at $T\neq T_{H}=0$. In the previous
paper \cite{non} we showed that nonextreme black holes with $T\neq T_{H}$
may exist. The reason in both cases is the same and I repeat it shortly. The
usual argument to reject the possibility $T\neq T_{H}$ relies on two facts:
1) this inequality makes the behavior of the stress-energy tensor of quantum
fields singular in the vicinity of the horizon, 2) in turn, such a behavior
of the stress-energy tensor is implied to inevitably destroy a regular
horizon by strong backreaction. Meanwhile, a new subtlety appears for
dilaton gravity as compared to the usual case. As, in addition to a metric
and quantum fields, there is one more object - dilaton, there exists the
possibility that divergencies in the stress-energy tensor are compensated
completely by gradients of a dilaton field to give a metric regular at the
horizon (at least, from outside). And for some models, provided the
conditions describe above are fulfilled, this situation is indeed realized.
Thus, dilaton gravity shares the point 1) with general relativity but the
condition 2) may break down.

It is worthwhile to note that, as follows from (\ref{bas}), the temperature
which determines the asymptotic value of the energy density at infinity, is
the function of the given coefficient $B$ which enters the form of the
action coefficient $F^{(0)}(\psi )$: $T=T_{0}\sqrt{1-B/\kappa }$, so the
solution under discussion has sense for $B\leq \kappa $ only. In particular,
for ''standard'' extreme holes with $T=0$ the coefficient $B=\kappa \neq 0$.
It follows from (\ref{exact}), (\ref{psi}), (\ref{psi}), that near the
horizon (i.e. in the limit $y\rightarrow -\infty $) the coefficient $V$ in
the action (\ref{clac}) behaves like $V\approx \frac{\omega ^{2}}{2}(\kappa
-B)$, so the condition $B\leq \kappa $ ensures the right sign of the term
with $(\nabla \phi )^{2}$.

The fact that for a given $B$ we obtain the fixed value of the temperature
is contrasted with that for nonextreme black holes with $T\neq T_{H}$ where
the temperature is not fixed by the form of the action but takes its value
within the whole range \cite{non}. This can be explained as follows. In the
extreme case we must achieve $T_{H}=0$ that entails fine tuning in the
asymptotic behavior of $\psi _{0}(y)$ that forces us to choose the
coefficient at the linear part of $F^{(0)}$ equal to $B$ exactly. Meanwhile,
in the nonextreme case, where the value of $T_{H}$ is not fixed beforehand,
the corresponding coefficient is bounded only by two conditions which
guarantee the existence of the horizon and the finiteness of the curvature.
These conditions result in two inequalities, so the constraint is less
restrictive (see \cite{non} for details).

It is worth recalling that Trivedi has found nonanalytical behavior of the
metric of extremal black holes near the horizon: $g\approx \alpha
_{1}(x-x_{h})^{2}+\alpha _{2}(x-x_{h})^{3+\delta }$ (see eq. (27) of \cite
{triv}). In our case corrections to the leading term of (\ref{g}) also may
be nonanalytical, depending on the properties of the function $F^{(0)}(\psi
_{0})$ (for the example (\ref{ex}) all is determined by the choice of the
function $\gamma (\psi _{0})$ ). Moreover, it is readily seen that in our
case the nonanaliticity may reveal itself in the leading term of $g$.
Indeed, let the function $g$ have the asymptotic 
\end{mathletters}
\begin{equation}
g\approx g_{0}(x-x_{h})^{\delta }  \label{geng}
\end{equation}
with $\delta >2$ that is compatible with the extremality condition $T_{H}=0$%
. This case can be handled in the same manner as for $\delta =2$. From (\ref
{bas}) and (\ref{sch}) it follows that now we must have $\,$near the horizon
the asymptotics 
\begin{equation}
\psi _{0}\approx 2y+\alpha \ln (-y)\text{,}  \label{genas}
\end{equation}
with $\alpha =\delta /(\delta -1)$ instead of (\ref{psi}). One more case
arises when $g\approx g_{0}e^{-x}$, so $y\sim e^{x}$, the horizon lies at $%
x=\infty $. Then we obtain the same formula (\ref{genas}) with $\alpha =1$.

From the physical viewpoint, the model considered in \cite{triv} represents
charged black holes in the dilaton theory motivated by the reduction from
four dimensions. The corresponding model is not exactly solvable either due
to the presence of the electromagnetic field or due to the form of the
action coefficients which do not fall into the class of exactly solvable
models \cite{zasl99}, \cite{kaz}. On the other hand, our model is exactly
solvable and deals with uncharged black holes. Their extremality is due to
quantum effects entirely.

The relevant physical object in dilaton gravity includes not only a metric
but both the metric and dilaton field. In the situations analyzed above the
coupling $F^{(0)}$ between curvature and dilaton describe inevitably
diverges at the horizon as seen from eq. (\ref{bas}) in the limit $%
y\rightarrow $ $-\infty $ corresponding to the horizon. Therefore, although
the curvature is finite on the horizon, the solution as the whole exhibit
singular behavior. Moreover, even the metric part of the solution can be
called ''regular'' only in the very restricted sense: the region inside the
horizon hardly has a physical meaning at all since an outer observer cannot
penetrate it because of strong divergencies in the stress-energy tensor on
the horizon surface, so the manifold is geodesically incomplete. All these
features tell us that in fact we deal with the class of objects which
occupies the intermediate positions between ''standard'' regular black holes
and naked singularities. In the previous paper we described the nonextreme
type of such objects, in the present paper we found its extreme version.

\section{summary}

We have proved that extreme black holes do exist in exactly solvable models
of dilaton gravity with quantum corrections taken into account. It turned
out that they are contained in general solutions of exactly solvable models.
The existence of extreme black holes depends strongly on the behavior of the
action coefficients near the horizon (see eq. (\ref{F})) and is insensitive
to their concrete form far from it where it is only assumed that their
dependence on the dilaton field ensures the absence of singularities.

We found solutions which shares features of both black holes and naked
singularities and represent ''singularities without singularities''. We
showed that quantum fields propagating in the background of extreme black
holes may nonzero temperature at infinity. In fact, this means that such
fields cannot be called Hawking radiation since the geometry does not
enforce the value of the temperature. These properties of solutions with $%
T\neq T_{H}$ are the same as in the nonextreme case \cite{non}. The
qualitatively new feature inherent to extreme black holes consists in that
the type of solutions we deal with is very sensitive to quantum effects:
taking the classical limit destroys completely our solutions with $T\neq
T_{H}=0$, so the solutions under discussion are due to quantum effects only.
(In contrast to it, solutions describing nonextreme black holes with $T\neq
T_{H}$ considered in \cite{non} have sense in the classical limit.)

The separate issue which deserves separate consideration is what value of
entropy should be ascribed to the objects we found.

I\ am grateful to Sergey Solodukhin for valuable comments.

% insert suggested PACS	numbers	in braces on next line

% body of paper	here

% now the references. delete or	change fake bibitem. delete next three
%   lines and directly read in your .bbl file if you use bibtex.

% Here is an example of	the general form of a figure:
% that you will	use with \ref{}	command	in the braces of the \label{} command.
%
% \begin{figure}
% \caption{}
% \label{}
% \end{figure}

% tables follow	here
%
% Here is an example of	the general form of a table:
% that you will	use with \ref{}	command	in the braces of the \label{} command.
% \begin{table}
% \caption{}
% \label{}
% \end{tabular}
% \end{table}


\end{document}